\documentstyle{article}
\newcommand{\K}{\mbox{\bf K}}
\newcommand{\R}{\mbox{\bf R}}
\newcommand{\C}{\mbox{\bf C}}
\newcommand{\Om}{{\bf\Omega}}
\newcommand{\cA}{{\cal A}}
\newcommand{\cB}{{\cal B}}
\newcommand{\cD}{{\cal D}}
\newcommand{\p}{\prime}
\newcommand{\e}{\mbox{\bf e}}
\newcommand{\M}{\mbox{M}}
\newcommand{\Ker}{\mbox{Ker}}
\begin{document}
\begin{flushright}
{\large\sf hep-th/9712217\\[1cm]}
\end{flushright}
\renewcommand{\thefootnote}{\fnsymbol{footnote}}
\begin{center}{\huge Physical fields and Clifford algebras II. Neutrino field}
\\[1.5cm]
{\large Vadim V. Varlamov\footnote[2]{E-mail: root@varlamov.kemerovo.su}}\\[0.3cm]
{\small\it Applied Mathematics, Siberian State Academy of Mining}\hspace{1mm}
{\&}\hspace{1mm}{\small\it Metallurgy,}\\
{\small\it Novokuznetsk, Russia}
\vspace{1.5cm}
\begin{abstract}
The neutrino field is considered in the framework of a complex Clifford
algebra $\C_{3}\cong\C_{2}\oplus\!\!\stackrel{\ast}{\C}_{2}$. 
The factor-algebras
${}^{\epsilon}\C_{2}$ and ${}^{\epsilon}\!\!\stackrel{\ast}{\C}_{2}$,
which are obtained by means of homomorphic mappings $\C_{3}\longrightarrow
\C_{2}$ and $\C_{3}\longrightarrow\stackrel{\ast}{\C}_{2}$, are identified
with the neutrino and antineutrino fields, respectively. In this framework
we have natural explanation for absence of right-handed neutrino and
left-handed antineutrino.
\end{abstract}
\end{center}
\renewcommand{\thefootnote}{\arabic{footnote}}
\thispagestyle{empty}
\newpage
\section{Fundamental automorphisms of Clifford algebra}
In Clifford algebra ${}^{l}\K_{n}$, where $\K$ is a field of characteristic
0 $(\K\!=\!\R,\;\K\!=\!\Om,\;\K\!=\!\C)$, there exists four fundamental
automorphisms \cite{1,2}:

1) An automorphism $\cA\longrightarrow\cA$.\\
This automorphism, obviously, be an identical automorphism of algebra
${}^{l}\K_{n}$, $\cA$ is arbitrary element of ${}^{l}\K_{n}$.

2) An automorphism $\cA\longrightarrow\cA^{\star}$.\\
In more details, for arbitrary element $\cA\in{}^{l}\K_{n}$ there exists
decomposition
$$\cA=\cA^{\p}+\cA^{\p\p},$$
where $\cA^{\p}$ is an element consisting of homogeneous odd elements, and
$\cA^{\p\p}$ is an element consisting of homogeneous even elements, 
respectively.
Then the automorphism $\cA\longrightarrow\cA^{\star}$ is that element
$\cA^{\p\p}$ is not changed, and element $\cA^{\p}$ is changed the sign:
$$\cA^{\star}=-\cA^{\p}+\cA^{\p\p}.$$
If $\cA$ is a homogeneous element, then
\begin{equation}\label{e1}
\cA^{\star}=(-1)^{k}\cA,
\end{equation}
where $k$ is a degree of element.

3) An anti-automorphism $\cA\longrightarrow\widetilde{\cA}$.\\
The anti-automorphism $\cA\longrightarrow\widetilde{\cA}$ be a reversion
of the element $\cA$, that is the substitution of each basis element
$\e_{i_{1}i_{2}\ldots i_{k}}\in\cA$ by element $\e_{i_{k}i_{k-1}\ldots i_{1}}$:
$$\e_{i_{k}i_{k-1}\ldots i_{1}}=(-1)^{\frac{k(k-1)}{2}}\e_{i_{1}i_{2}\ldots
i_{k}}.$$
Therefore, for any $\cA\in{}^{l}\K_{n}$ we have
\begin{equation}\label{e2}
\widetilde{\cA}=(-1)^{\frac{k(k-1)}{2}}\cA.
\end{equation}

4) An anti-automorphism $\cA\longrightarrow\widetilde{\cA^{\star}}$.\\
This anti-automorphism be a composition of the anti-automorphism
$\cA\longrightarrow\widetilde{\cA}$ with the automorphism 
$\cA\longrightarrow\cA^{\star}$. In the case of homogeneous element from
formulae (\ref{e1}) and (\ref{e2}) follows
\begin{equation}\label{e3}
\widetilde{\cA^{\star}}=(-1)^{\frac{k(k+1)}{2}}\cA.
\end{equation}
It is obvious that\hspace{1mm} $\widetilde{\!\!\widetilde{\cA}}=\cA,\;
(\cA^{\star})^{\star}=
\cA,$ and $\widetilde{(\widetilde{\cA^{\star}})^{\star}}=\cA$.

For example consider the anti-automorphism $\cA\longrightarrow\widetilde{\cA}$
for electromagnetic field. It is known \cite{3,4} that
the components of electric and magnetic
vectors may be represented by the following element of algebra $\R_{3}$:
\begin{equation}\label{e4}
\cA=E^{1}\e_{1}+E^{2}\e_{2}+E^{3}\e_{3}+H^{1}\e_{23}+H^{2}\e_{31}+H^{3}\e_{12}.
\end{equation}
Since the maximal basis element $\omega=\e_{123}\;(\omega^{2}=-1)$ is
belong to a center of algebra $\R_{3}$ the element (\ref{e4}) (by force of
identity $\R_{3}=\C_{2}$ \cite{4}) may be rewritten as
\begin{equation}\label{e5}
\cA=\underbrace{(E^{1}+\omega H^{1})}_{F^{1}}\e_{1}+
\underbrace{(E^{2}+\omega H^{2})}_{F^{2}}\e_{2}+
\underbrace{(E^{3}+\omega H^{3})}_{F^{3}}\e_{3}.
\end{equation}
Further, under action of anti-automorphism $\cA\longrightarrow\widetilde{\cA}$
the elements (\ref{e4}) and (\ref{e5}) are adopt the following form
(by force of (\ref{e2}))
\begin{equation}\label{e6}
\widetilde{\cA}=E^{1}\e_{1}+E^{2}\e_{2}+E^{3}\e_{3}-H^{1}\e_{23}-H^{2}\e_{31}-
H^{3}\e_{12},
\end{equation}
\begin{equation}\label{e7}
\widetilde{\cA}=\underbrace{(E^{1}-\omega H^{1})}_{\stackrel{\ast}{F^{1}}}\e_1
+\underbrace{(E^{2}-\omega H^{2})}_{\stackrel{\ast}{F^{2}}}\e_{2}
+\underbrace{(E^{3}-\omega H^{3})}_{\stackrel{\ast}{F^{3}}}\e_{3}.
\end{equation}

It is easy to see that elements (\ref{e5}) and (\ref{e7}) are isomorphic to
the general elements of algebras $\C_{2}$ and $\stackrel{\ast}{\C}_{2}$:
$\cA=F^{0}\e_{0}+F^{1}\e_{1}+F^{2}\e_{2}+F^{3}\e_{12}$ and 
\raisebox{0.3ex}{$\stackrel{\hspace{1mm}\ast}{\cA}$}$\;=\;\stackrel{\ast}{F^{0}}
\!\e_{0}+
\stackrel{\ast}{F^{1}}~\!\!\e_{1}+\stackrel{\ast}{F^{2}}~\!\!\e_{2}+\newline
\stackrel{\ast}{F^{3}}\!\e_{12}$, where $F^{0}=\stackrel{\ast}{F^{0}}=0$.
The algebras $\C_{2}$ and $\stackrel{\hspace{-1mm}\ast}{\C_{2}}$ are 
described the photon
fields with left-handed and right-handed polarization, respectively.
Therefore, under action of anti-automorphism $\cA\longrightarrow
\widetilde{\cA}$ the photon field with left-handed polarization turn into
the photon field with right-handed polarization and back.

The other example, also impotent in physical applications, we have for the
anti-automorphism $\cA\longrightarrow\widetilde{\cA^{\star}}$ which may be
connected with a charge conjugation \cite{2}. 
\section{Clifford algebras with odd dimensionality and \protect\newline 
spinor representations}
Later on we shall restricted by the case of a field $\K\!=\!\C$, because
this case is a most interesting for physics. Consider an algebra $\C_{3}$,
the general element of which has a following form:
$$\cA=a^{0}\e_{0}+\sum^{3}_{i=1}a^{i}\e_{i}+\sum^{3}_{i=1}\sum^{3}_{j=1}
a^{ij}\e_{ij}+a^{123}\e_{123}.$$
Here the volume element $\omega=\e_{123}\;(\omega^{2}=-1)$ is belong to a
center of algebra $\C_{3}$ (that is commutes with all elements of $\C_{3}$).
Let
\begin{equation}\label{e8}
\lambda_{+}=\frac{1+i\omega}{2},\quad\lambda_{-}=\frac{1-i\omega}{2}
\end{equation}
the mutually orthogonal idempotents of algebra $\C_{3}$ are satisfying to
conditions
$$
\begin{array}{rl}
\lambda_{+}+\lambda_{-}=1, & \lambda_{+}\lambda_{-}=0, \\
\lambda^{2}_{+}=\lambda_{+}, & \lambda^{2}_{-}=\lambda_{-}.
\end{array}$$
At this the idempotents $\lambda_{+}$ and $\lambda_{-}$ together with volume
element $\omega$ are commutes with all elements of $\C_{3}$ whose general
element now may be represented as
$$\cA=(\lambda_{+}+\lambda_{-})\cA=\lambda_{+}\cA+\lambda_{-}\cA.$$
 
Consider the elements of the form $\lambda_{-}\cA$. Choosing by turns
$\cA=\e_{0},\,\e_{1},\,\e_{2},\,\e_{3}$ we obtain
\begin{equation}\label{e9}
\frac{1}{2}(1-i\e_{123}),\quad\frac{1}{2}(\e_{1}-i\e_{23}),\quad
\frac{1}{2}(\e_{2}-i\e_{31}),\quad\frac{1}{2}(\e_{3}-i\e_{12}).
\end{equation}
Analogously, for the elements of the form $\lambda_{+}\cA$
\begin{equation}\label{e10}
\frac{1}{2}(1+i\e_{123}),\quad\frac{1}{2}(\e_{1}+i\e_{23}),\quad
\frac{1}{2}(\e_{2}+i\e_{31}),\quad\frac{1}{2}(\e_{3}+i\e_{12}).
\end{equation}
It is obvious that the same result we obtain if $\cA=\e_{23},\,\e_{31},\,
\e_{12},\,\e_{123}$. It is easy to see that elements $\lambda_{+}\cA$ are
make a subalgebra of $\C_{3}$ since
$$\lambda_{+}\cA+\lambda_{-}\cA^{\p}=\lambda_{+}(\cA+\cA^{\p}),\qquad
\lambda_{+}\cA\lambda_{+}\cA^{\p}=\lambda_{+}\cA\cA^{\p},$$
$$a\lambda_{+}\cA=\lambda_{+}a\cA.$$
The elements of the form $\lambda_{-}\cA$ also are make a subalgebra of
$\C_{3}$. The subalgebras $\{\lambda_{-}\cA\}$ and $\{\lambda_{+}\cA\}$ are
isomorphic (by force of identity $i\equiv\omega$) to algebras $\C_{2}$
and $\stackrel{\ast}{\C}_{2}$, respectively. Thus, the algebra $\C_{3}$ is
decomposed into a direct sum of two subalgebras $\{\lambda_{-}\cA\}\cong\C_{2}$
and $\{\lambda_{+}\cA\}\cong\stackrel{\ast}{\C}_{2}$ (the product of elements
from different subalgebras is equal to zero since $\lambda_{+}\lambda_{-}=0$).
Analogously, for the algebra $\C_{5}$ we have the same decomposition. By
means of idempotents $\lambda_{+}=\frac{1+\omega}{2}$ and $\lambda_{-}=
\frac{1-\omega}{2}\;(\omega=\e_{12345},\,\omega^{2}=1)$ the algebra $\C_{5}$
is decomposed into a direct sum of two subalgebras $\{\lambda_{+}\cA\}\cong
\C_{4}$ and $\{\lambda_{-}\cA\}\cong\stackrel{\ast}{\C}_{4}$. In general
\begin{equation}\label{e11}
{\renewcommand{\arraystretch}{1.3}
\begin{array}{ccl}
\C_{4m^{\p}+1}&\cong&\C_{4m^{\p}}\oplus\stackrel{\ast}{\C}_{4m^{\p}},\\
\C_{4m^{\p}-1}&\cong&\C_{4m^{\p}-2}\oplus\stackrel{\ast}{\C}_{4m^{\p}-2},
\end{array}}\end{equation}
$$m^{\p}=1,2\ldots .$$
At this the mutually orthogonal idempotents have the following form
\begin{eqnarray}
\lambda_{+}&=&\frac{1+i\omega}{2},\qquad\lambda_{-}\;=\;\frac{1-i\omega}{2},
\qquad\mbox{if}\;\;\omega^{2}\;=\;-1,\nonumber\\
\lambda_{+}&=&\frac{1+\omega}{2},\qquad\phantom{i}\lambda_{-}\;=\;
\frac{1-\omega}{2},\qquad\phantom{i}
\mbox{if}\;\;\omega^{2}\;=\;1.\nonumber
\end{eqnarray}

It is well-known \cite{2} that Clifford algebras $\C_{n}$ with even 
dimensionality
($n=2\nu$) are isomorphic to matrix algebras of order $2^{\nu}$ 
($\M_{2^{\nu}}(\C)$). Hence it immediately follows that linear (matrix or
spinor) representations of Clifford algebras with any dimensionality over
a field $\K\!=\!\C$ are defined by the following isomorphisms:
\begin{eqnarray}
\C_{4m^{\p}}&\cong&\M_{2^{2m^{\p}}}(\C),\nonumber \\
\C_{4m+2}&\cong&\M_{2^{2m+1}}(\C),\nonumber \\
\C_{4m^{\p}+1}&\cong&\M_{2^{2m^{\p}}}(\C)\oplus\M_{2^{2m^{\p}}}(\stackrel{\ast}
{\C}),\nonumber \\
\C_{4m^{\p}-1}&\cong&\M_{2^{2m^{\p}-1}}(\C)\oplus\M_{2^{2m^{\p}-1}}
(\stackrel{\ast}{\C}),\nonumber
\end{eqnarray}
$$m=0,1,\ldots ,$$
$$m^{\p}=1,2,\ldots .$$
\section{A homomorphism $\C_{n+1}\longrightarrow\C_{n}$}\begin{sloppypar}
From the latest isomorphisms follows that the linear representations of
algebras $\C_{n}$ with $n$ is odd are isomorphic to the direct sums of matrix
algebras $\M_{2^{2m^{\p}}}(\C)\oplus\M_{2^{2m^{\p}}}(\C)$ and 
$\M_{2^{2m^{\p}-1}}(\C)\oplus\M_{2^{2m^{\p}-1}}(\C)$, respectively.
However, there exists a homomorphism between Clifford algebras
$\C_{4m^{\p}+1},\;\C_{4m^{\p}-1}$ and matrix algebras $\M_{2^{2m^{\p}}}(\C),\;
\M_{2^{2m^{\p}-1}}(\C)$. Consider this homomorphism in details.
First of all, the volume elements $\omega=\e_{12\ldots 4m^{\p}+1}$ and
$\omega=\e_{12\ldots 4m^{\p}-1}$ are belong to the centers of algebras
$\C_{4m^{\p}+1}$ and $\C_{4m^{\p}-1}$, and therefore are commutes with all
basis elements of these algebras. Further, since $\omega^{2}=
\e^{2}_{12\ldots n}=(-1)^{\frac{n(n-1)}{2}}$, where $n=4m^{\p}+1$ or
$n=4m^{\p}-1$, then denoting $\varepsilon=\pm i^{\frac{n(n-1)}{2}}$ we shall
have always $(\varepsilon\omega)^{2}=1$. Recalling that the each Clifford
algebra $\C_{n}$ is associate with a complex vector space $C_{n}$ \cite{5,6} 
we see
that the basis vectors $\{e_{1},e_{2},\ldots,e_{n}\}$ are spanned a 
subspace $C_{n}\subset C_{n+1}$. Thus, the algebra $\C_{n}$ in $C_{n}$ be
a subalgebra of $\C_{n+1}$ and consist of the elements which are not contain
the symbol $\e_{n+1}$ (in our case $\e_{4m^{\p}+1}$ and $\e_{4m^{\p}-1}$).
A decomposition of the each element $\cA\in\C_{n+1}$ may be written in the
following form\end{sloppypar}
$$\cA=\cA^{1}+\cA^{0},$$
where $\cA^{0}$ be a set whose elements are contain the unit $\e_{n+1}$, and
$\cA^{1}$ be a set whose elements are not contain $\e_{n+1}$, therefore
$\cA^{1}\in\C_{n}$. If multiply $\cA^{0}$ by $\varepsilon\omega$, then the
units $\e_{n+1}$ are mutually annihilate, therefore $\varepsilon\omega
\cA^{0}\in\C_{n}$. Denoting $\cA^{2}=\varepsilon\omega\cA^{0}$ we obtain
$$\cA=\cA^{1}+\varepsilon\omega\cA^{2},$$
where $\cA^{1},\,\cA^{2}\in\C_{n}$. Define now a homomorphism
$\epsilon:\;\C_{n+1}\longrightarrow\C_{n}$, the action of which satisfy to
the following rule
\begin{equation}\label{e12}
\epsilon:\;\cA^{1}+\varepsilon\omega\cA^{2}\longrightarrow\cA^{1}+\cA^{2}.
\end{equation}
Obviously, at this the all operations (addition, multiplication, and
multiplication by number) are remained. Indeed, let
$$\cA=\cA^{1}+\varepsilon\omega\cA^{2},\quad
\cB=\cB^{1}+\varepsilon\omega\cB^{2},$$  
then for multiplication by force of $(\varepsilon\omega)^{2}=1$ and 
commutativity $\omega$ with all elements we have
$$\cA\cB=(\cA^{1}\cB^{1}+\cA^{2}\cB^{2})+\varepsilon\omega(\cA^{1}\cB^{2}+
\cA^{2}\cB^{1})\stackrel{\epsilon}{\longrightarrow}(\cA^{1}\cB^{1}+
\cA^{2}\cB^{2})+(\cA^{1}\cB^{2}+\cA^{2}\cB^{1})=$$
$$=(\cA^{1}+\cA^{2})(\cB^{1}+\cB^{2}).$$
Thus, the image of product is equal to the product of multiplier images in
the same order.

In the particular case when $\cA=\varepsilon\omega$ we have $\cA^{1}=0$ and
$\cA^{2}=1$, therefore
$$\varepsilon\omega\longrightarrow 1.$$
This way, a kernel of homomorphism $\epsilon$ consist of the all elements
of form $\cA^{1}-\varepsilon\omega\cA^{1}$. Under action of homomorphism
$\epsilon$ the elements $\cA^{1}-\varepsilon\omega\cA^{1}$ are mapped into
zero. Clearly, that $\Ker\,\epsilon=\{\cA^{1}-\varepsilon\omega\cA^{1}\}$ be
also a subalgebra of $\C_{n+1}$. Moreover, the kernel of homomorphism 
$\epsilon$ be a bilateral ideal of algebra $\C_{n+1}$. Therefore, the algebra
$\C_{n}$, which we obtain in the result of mapping $\epsilon:\;\C_{n+1}
\longrightarrow\C_{n}$, be a factor-algebra by the ideal $\Ker\,\epsilon=
\{\cA^{1}-\varepsilon\omega\cA^{1}\}$:
$$\C_{n}\cong\C_{n+1}/\Ker\,\epsilon=\{\cA^{1}+\varepsilon\omega\cA^{2}\}/
\{\cA^{1}-\varepsilon\omega\cA^{1}\}.$$

Recalling that algebra $\C_{n}$, where $n$ is even, is isomorphic to a matrix
algebra $\M_{2^{n/2}}(\C)$ we obtain by force of $\epsilon:\;\C_{n+1}
\longrightarrow\C_{n}\subset\C_{n+1}$ that algebra $\C_{n+1}$ is isomorphic
to a matrix algebra $\M_{2^{n/2}}(\C)$. For more details
\begin{eqnarray}
\C_{4m^{\p}+1}&\longrightarrow&\M_{2^{2m^{\p}}}(\C),\nonumber \\
\C_{4m^{\p}-1}&\longrightarrow&\M_{2^{2m^{\p}-1}}(\C).\nonumber
\end{eqnarray}
 
Consider now the fundamental automorphisms $\cA\longrightarrow\widetilde{\cA},
\;\cA\longrightarrow\cA^{\star},\;\cA\longrightarrow\widetilde{\cA^{\star}}$
are defined in $\C_{4m^{\p}+1}$ or $\C_{4m^{\p}-1}$. It is interest to us
what form these automorphisms are adopted after homomorphic mapping
$\epsilon:\;\C_{n+1}\longrightarrow\C_{n}\subset\C_{n+1}$. First of all,
in the case of anti-automorphism $\cA\longrightarrow\widetilde{\cA}$ the
elements $\cA,\,\cB,\,\ldots\,\in\C_{n+1}$, which are mapped into one element
$\cD\in\C_{n}$ (the kernel of homomorphism $\epsilon$ if $\cD=0$) after
transformation $\cA\longrightarrow\widetilde{\cA}$ are must transit to the
elements $\widetilde{\cA},\,\widetilde{\cB},\,\ldots\,\in\C_{n+1}$ 
which are also
mapped into one element $\widetilde{\cD}\in\C_{n}$. Otherwise the
transformation $\cA\longrightarrow\widetilde{\cA}$ not be one-to-one. In
particular, it is necessary that $\widetilde{\varepsilon\omega}=\varepsilon
\omega$, since $1$ and element $\varepsilon\omega$ under action of 
homomorphism $\epsilon$ are equally mapped into unit, then $\widetilde{1}$
and $\widetilde{\varepsilon\omega}$ are also mapped into one element, but
$\widetilde{1}\longrightarrow 1$, and $\widetilde{\varepsilon\omega}
\longrightarrow\pm 1$ (by force of (\ref{e2})), therefore we must assume
\begin{equation}\label{e13}
\widetilde{\varepsilon\omega}=\varepsilon\omega.
\end{equation}
The condition (\ref{e13}) be sufficient for transition of anti-automorphism
$\cA\longrightarrow\widetilde{\cA}$ from $\C_{n+1}$ to $\C_{n}$. Indeed,
in this case we have
$$\cA^{1}-\cA^{1}\varepsilon\omega\;\longrightarrow\;\widetilde{\cA^{1}}-
\widetilde{\varepsilon\omega}\widetilde{\cA^{1}}=\widetilde{\cA^{1}}-
\varepsilon\omega\widetilde{\cA^{1}}.$$
Therefore, the elements of the form $\cA^{1}-\cA^{1}\varepsilon\omega$, which
are belong to a kernel of homomorphism $\epsilon$, at the transformation
$\cA\longrightarrow\widetilde{\cA}$ are transit to the elements of the same
form.

The analogous conditions we have for other fundamental automorphisms.
However, for automorphism $\cA\longrightarrow\cA^{\star}$ a condition
$(\varepsilon\omega)^{\star}=\varepsilon\omega$ is not execute, since
$\omega$ is odd, and in accordance with (\ref{e1}) we have
$$\omega^{\star}=-\omega.$$
Therefore, the automorphism $\cA\longrightarrow\cA^{\star}$ never transfer
from $\C_{n+1}$ to $\C_{n}$. 

Further, in general case we have for a field $\K\!=\!\C$ two homomorphisms:
\begin{eqnarray}
\C_{4m^{\p}+1}&\longrightarrow&\C_{4m^{\p}},\nonumber \\
\C_{4m^{\p}-1}&\longrightarrow&\C_{4m^{\p}-2}.\nonumber
\end{eqnarray}
The multiplier $\varepsilon=\pm i^{\frac{n(n-1)}{2}}$ (here $n$ is equal
to $4m^{\p}+1$ or $4m^{\p}-1$) in this case has the following values
$$\varepsilon=\left\{   
\begin{array}{rl}
\pm 1, & \mbox{if}\; n=4m^{\p}+1,\\
\pm i, & \mbox{if}\; n=4m^{\p}-1;
\end{array}\right.$$
Thus, for anti-automorphism $\cA\longrightarrow\widetilde{\cA}$ in accordance
with (\ref{e2}) we have
\begin{eqnarray}
\widetilde{\omega}&=&\widetilde{\e}_{12\ldots 4m^{\p}+1}=(-1)^{(4m^{\p}+1)
4m^{\p}}\omega=\omega, \\
\widetilde{i\omega}&=&i\widetilde{\e}_{12\ldots 4m^{\p}-1}=(-1)^{(4m^{\p}-1)
(2m^{\p}-1)}i\omega=-i\omega.\label{e15}
\end{eqnarray}
Therefore, by force of condition (\ref{e13}) the anti-automorphism
$\cA\longrightarrow\widetilde{\cA}$ is transfer only in the case
$$\C_{4m^{\p}+1}\;\longrightarrow\;\C_{4m^{\p}}.$$

Consider now the anti-automorphism $\cA\longrightarrow\widetilde{\cA^{\star}}$.
Obviously, for transfer of $\cA\longrightarrow\widetilde{\cA^{\star}}$ from
$\C_{n+1}$ to $\C_{n}$ it is necessary that the following condition be
executed:
\begin{equation}\label{e16}
\widetilde{(\varepsilon\omega)^{\star}}=\varepsilon\omega.
\end{equation}
It is easy to see that by force of (\ref{e15}) this condition be executed
only in the case
$$\C_{4m^{\p}-1}\;\longrightarrow\;\C_{4m^{\p}-2},$$
since
$$\widetilde{(\varepsilon\omega)^{\star}}=\varepsilon\widetilde{\e^{\star}}
_{12\ldots 4m^{\p}-1}=(-1)^{(4m^{\p}-1)(2m^{\p}-1)}\varepsilon\omega^{\star}=
-\varepsilon\omega^{\star}=\varepsilon\omega.$$

Finally, from $\C_{n+1}$ to $\C_{n}$ are transfer the following fundamental
automorphisms:
\begin{eqnarray}
\cA&\longrightarrow&\widetilde{\cA}\quad\phantom{a}\mbox{at}\quad\C_{4m^{\p}+1}
\longrightarrow\C_{4m^{\p}},\label{e16'} \\
\cA&\longrightarrow&\widetilde{\cA^{\star}}\quad\mbox{at}\quad\C_{4m^{\p}-1}
\longrightarrow\C_{4m^{\p}-2}.\label{e16''}
\end{eqnarray}
\newpage
\section{Neutrino field}
The main goal of present paper be a consideration of neutrino field in the
framework of algebra $\C_{3}$, which be a simplest algebra with odd
dimensionality. At this the central role in this consideration played a
homomorphism $\epsilon:\;\C_{n+1}\longrightarrow\C_{n}$. First of all,
in accordance with (\ref{e11}) the algebra $\C_{3}$ is decomposed into a
direct sum of two subalgebras $\C_{2}$ and $\stackrel{\ast}{\C}_{2}$. This
decomposition may be represented by a following diagram:
$$
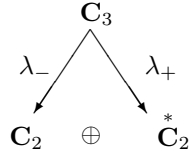
\unitlength=0.5mm
\begin{picture}(70,50)
\put(35,40){\vector(2,-3){15}}
\put(35,40){\vector(-2,-3){15}}
\put(32.25,42){$\C_{3}$}
\put(16,28){$\lambda_{-}$}
\put(49.5,28){$\lambda_{+}$}
\put(13.5,9.20){$\C_{2}$}
\put(52.75,9){$\stackrel{\ast}{\C}_{2}$}
\put(32.5,10){$\oplus$}
\end{picture}$$
Here we identify the idempotents $\lambda_{-},\,\lambda_{+}$ with the
helicity projection operators. Further, after homomorphic mappings
$\C_{3}\longrightarrow\C_{2}$ and $\C_{3}\longrightarrow\stackrel{\ast}
{\C}_{2}$ we obtain in result the factor-algebras ${}^{\epsilon}\C_{2}$ and
${}^{\epsilon}\!\!\stackrel{\ast}{\C}_{2}$ (we introduce here the sign of 
homomorphism $\epsilon$ for difference of neutrino fields from the photon
fields with left-handed and right-handed polarizations, which are described 
by the algebras $\C_{2}$ and $\stackrel{\ast}{\C}_{2}$). Later on we shall
connected the factor-algebras ${}^{\epsilon}\C_{2}$ and ${}^{\epsilon}\!\!
\stackrel{\ast}{\C}_{2}$ with the neutrino and antineutrino fields,
respectively. Indeed, the factor-algebra ${}^{\epsilon}\C_{2}$ (like to
$\C_{2}$) is isomorphic to a matrix algebra $\M_{2}(\C)$ which is represent
the algebra of linear operators in a 2-dimensional complex space
$S_{2}$\footnote{It is obvious that this space is homeomorphic to extended
complex plane. Moreover, the space $S_{2}$ may be identified with an absolute
(a totality of infinitely distant points) of Lobatchevskii space 
${}^{1}S_{3}$ \cite{7}, the motion group of which is a Lorentz group 
(at this the
space ${}^{1}S_{3}$ be an absolute of Minkowski space-time ${}^{1}R_{4}$
\cite{8}).}
(so-called a spinor space). Analogously, by force of isomorphism
${}^{\epsilon}\!\!\stackrel{\ast}{\C}_{2}\cong\M_{2}(\stackrel{\ast}{\C})$ 
we have a space $\stackrel{\ast}{S}_{2}$ (a co-spinor space). The linear
transformations of the vectors (spinors) of these spaces are defined by the
following expressions:
$$\vspace{1mm}{\renewcommand{\arraystretch}{1.7}
\begin{array}{ccc}
{\xi^{1}}^{\p}&=&\alpha\xi^{1}+\beta\xi^{2}, \\
{\xi^{2}}^{\p}&=&\gamma\xi^{1}+\delta\xi^{2},
\end{array}}\qquad
\begin{array}{ccc}
{\stackrel{\ast}{\xi^{1}}}^{\p}&=&\stackrel{\ast}{\alpha}\stackrel{\ast}
{\xi^{1}}+\stackrel{\ast}{\beta}\stackrel{\ast}{\xi^{2}},\\
{\stackrel{\ast}{\xi^{2}}}^{\p}&=&\stackrel{\ast}{\gamma}\stackrel{\ast}
{\xi^{1}}+\stackrel{\ast}{\delta}\stackrel{\ast}{\xi^{2}},
\end{array}$$
$$\sigma=\left({\renewcommand{\arraystretch}{1.2}
\begin{array}{cc}
\alpha & \beta \\
\gamma & \delta
\end{array}}\right),\qquad\stackrel{\ast}{\sigma}=\left(
\begin{array}{cc}
\stackrel{\ast}{\alpha} & \stackrel{\ast}{\beta}\\
\stackrel{\ast}{\gamma} & \stackrel{\ast}{\delta}
\end{array}\right).$$
It is well-known that a group of all complex matrices of second order is
isomorphic to a group $SL(2;\C)$, which be a double-meaning representation
of the Lorentz group. We come to van der Waerden 2-spinor formalism 
\cite{9,10,11} if
suppose $\stackrel{\ast}{\xi^{1}}\sim\xi^{\dot{1}},\,\stackrel{\ast}{\xi^{2}}
\sim\xi^{\dot{2}}$. At this the spaces of un-dotted and dotted spinors
$S_{2}$ and $\dot{S}_{2}$ be respectively the spaces of un-dotted and dotted
representations of the own Lorentz group. In this case an any spin-tensor
may be represented as $a^{n_{1}\ldots n_{r}}_{\dot{m}_{1}\ldots\dot{m}_{s}}=
\xi^{n_{1}}\cdots\xi^{n_{r}}\stackrel{\ast}{\xi}_{m_{1}}\cdots\stackrel{\ast}
{\xi}_{m_{s}}\;(n_{i},m_{j}=1,2)$. Let
$$\xi^{\lambda}=\left(
\begin{array}{c}
\xi^{1} \\
\xi^{2}
\end{array}\right),\quad\eta_{\dot{\mu}}=\left(
\begin{array}{c}
\eta_{\dot{1}} \\
\eta_{\dot{2}}
\end{array}\right)$$
are two-component spinors, obviously, be the vectors of the spaces $S_{2}$
and $\dot{S}_{2}$, and are satisfying to conditions
\begin{equation}\label{e17}
\xi_{1}=\xi^{2},\quad\xi_{2}=-\xi^{1},\quad\eta^{\dot{1}}=-\eta_{\dot{2}},
\quad\eta^{\dot{2}}=\eta_{\dot{1}}.
\end{equation}
Further, let
$$\left(
\begin{array}{cc}
\partial_{1\dot{1}} & \partial_{1\dot{2}} \\
\partial_{2\dot{1}} & \partial_{2\dot{2}}
\end{array}\right)=\left(
\begin{array}{cc}
\partial_{0}-\partial_{3} & \partial_{1}-i\partial_{2} \\
\partial_{1}+i\partial_{2} & \partial_{0}+\partial_{3}
\end{array}\right),$$
$$\left(
\begin{array}{cc}
\partial^{1\dot{1}} & \partial^{2\dot{1}} \\
\partial^{1\dot{2}} & \partial^{2\dot{2}}
\end{array}\right)=\left(
\begin{array}{cc}
\partial_{0}+\partial_{3} & \partial_{1}+i\partial_{2} \\
\partial_{1}-i\partial_{2} & \partial_{0}-\partial_{3}
\end{array}\right).$$
are matrices of the symmetric spin-tensors $\partial_{\lambda\dot{\mu}}=
\partial_{\dot{\mu}\lambda},\;\partial^{\lambda\dot{\mu}}=\partial^{\dot{\mu}
\lambda}$. The matrix of spin-tensor $\partial^{\lambda\dot{\mu}}$ is
obtained from the matrix of $\partial_{\lambda\dot{\mu}}$ by means of
conjugation and formulae (\ref{e17}).

Compose now the equations $\partial_{\lambda\dot{\mu}}\xi^{\lambda}=0,\;
\partial^{\lambda\dot{\mu}}\eta_{\dot{\mu}}=0$:
\begin{equation}\label{e18}{\renewcommand{\arraystretch}{1.4}
\begin{array}{ccc}
\partial_{1\dot{1}}\xi^{1}+\partial_{1\dot{2}}\xi^{2}&=&0,\\
\partial_{2\dot{1}}\xi^{1}+\partial_{2\dot{2}}\xi^{2}&=&0,\\
\partial^{1\dot{1}}\eta_{\dot{1}}+\partial^{2\dot{1}}\eta_{\dot{2}}&=&0,\\
\partial^{1\dot{2}}\eta_{\dot{1}}+\partial^{2\dot{2}}\eta_{\dot{2}}&=&0.
\end{array}}\end{equation}      
It is well-known that equations $\partial_{\lambda\dot{\mu}}\xi^{\lambda}=0,\;
\partial^{\lambda\dot{\mu}}\eta_{\dot{\mu}}=0$ be the Weyl equations for
neutrino and antineutrino fields, respectively. In connection with this
consider one interesting correlation between Weyl equations and Maxwell
equations in vacuum. Maxwell equations in the spinor form may be written as
\begin{eqnarray}
\partial^{\dot{\mu}}_{\lambda}f^{\lambda}_{\rho}&=&0,\nonumber \\
\partial^{\dot{\mu}}_{\lambda}f^{\lambda}_{\rho}&=&s^{\dot{\mu}}_{\rho}.
\nonumber \end{eqnarray}
Or, in the case of $s^{\dot{\mu}}_{\rho}=0$:
\begin{eqnarray}
\partial^{\dot{1}}_{2}f^{1}_{2}+\partial^{\dot{2}}_{2}f^{2}_{2}&=&0,\nonumber\\
\partial^{\dot{1}}_{1}f^{1}_{2}+\partial^{\dot{2}}_{1}f^{2}_{2}&=&0,\nonumber\\
\partial^{\dot{1}}_{1}f^{1}_{1}+\partial^{\dot{2}}_{1}f^{2}_{1}&=&0,\nonumber\\
\partial^{\dot{1}}_{2}f^{1}_{1}+\partial^{\dot{2}}_{2}f^{2}_{1}&=&0.\nonumber
\end{eqnarray}
Using (\ref{e17}) the latest system may be rewritten in the form
\begin{eqnarray}
\partial_{2\dot{1}}f_{12}+\partial_{2\dot{2}}f_{22}&=&0,\nonumber\\
\partial_{1\dot{1}}f_{12}+\partial_{1\dot{2}}f_{22}&=&0,\nonumber\\
\partial_{1\dot{1}}f_{11}+\partial_{1\dot{2}}f_{12}&=&0,\nonumber\\
\partial_{2\dot{1}}f_{11}+\partial_{2\dot{2}}f_{12}&=&0,\nonumber
\end{eqnarray}
Raising the indexes in the first two equations of the latest system and
conjugate these equations (using the property $\dot{\!\dot{\xi}}=\xi$)
we obtain
\begin{equation}\label{e19}{\renewcommand{\arraystretch}{1.4}
\begin{array}{ccc}
\partial^{1\dot{1}}f^{\dot{1}\dot{1}}+\partial^{2\dot{1}}f^{\dot{1}\dot{2}}
&=&0,\nonumber\\
\partial^{1\dot{2}}f^{\dot{1}\dot{1}}+\partial^{2\dot{2}}f^{\dot{1}\dot{2}}
&=&0,\nonumber\\
\partial_{1\dot{1}}f_{11}+\partial_{1\dot{2}}f_{12}&=&0,\nonumber\\
\partial_{2\dot{1}}f_{11}+\partial_{2\dot{2}}f_{12}&=&0.
\end{array}}\end{equation}
It is easy to see that system (\ref{e18}) is coincide with the system
(\ref{e19}) if suppose
$$\xi^{\lambda}\sim\left(
{\renewcommand{\arraystretch}{1.3}
\begin{array}{c}
f_{11}\\
f_{12}
\end{array}}\right)=\left(
{\renewcommand{\arraystretch}{1.3}
\begin{array}{c}
F^{3}\\
F^{1}+iF^{2}
\end{array}}\right),\quad\eta_{\dot{\mu}}\sim\left(
\begin{array}{c}
f^{\dot{1}\dot{1}}\\
f^{\dot{1}\dot{2}}
\end{array}\right)=\left(
\begin{array}{c}
\stackrel{\ast}{F^{3}}\\
\stackrel{\ast}{F^{1}}-i\stackrel{\ast}{F^{2}}
\end{array}\right).$$

Thus, we assume that neutrino and antineutrino fields are described by the
factor-algebras ${}^{\epsilon}\C_{2}$ and ${}^{\epsilon}\!\!\stackrel{\ast}
{\C}_{2}$. It is well-known that in Nature there exists only left-handed
neutrino and right-handed antineutrino, and no there exists right-handed
neutrino and left-handed antineutrino. In other words, the transformations
of neutrino (antineutrino) field are not contain spatial reflections, 
unlike to the photon fields, for which we have both left-handed and
right-handed photons are described by the algebras $\C_{2}$ and
$\stackrel{\ast}{\C}_{2}$, and are transformed into each other under action
of anti-automorphism $\cA\longrightarrow\widetilde{\cA}$. The absence of
spatial reflections in the case of neutrino field has a natural explanation
in the framework of algebra $\C_{3}$. Indeed, in accordance with (\ref{e16''})
at the homomorphic mappings $\C_{3}\longrightarrow\C_{2}$ and
$\C_{3}\longrightarrow\stackrel{\ast}{\C}_{2}$ the anti-automorphism
$\cA\longrightarrow\widetilde{\cA}$ is not transfer. Thus, for the
factor-algebras ${}^{\epsilon}\C_{2}$ and ${}^{\epsilon}\!\!\stackrel{\ast}
{\C}_{2}$, which are obtained in the result of these mappings, the 
anti-automorphism $\cA\longrightarrow\widetilde{\cA}$ is not defined
(that is the factor-algebras ${}^{\epsilon}\C_{2}$ and ${}^{\epsilon}\!\!
\stackrel{\ast}{\C}_{2}$ are not transformed into each other under action
of $\cA\longrightarrow\widetilde{\cA}$). Therefore, the neutrino and
antineutrino fields, which are described by ${}^{\epsilon}\C_{2}$ and
${}^{\epsilon}\!\!\stackrel{\ast}{\C}_{2}$, are possess the fixed
helicities (that is the transformations of these fields are not contain the
spatial reflections). However, in accordance with (\ref{e16''}) the
factor-algebras ${}^{\epsilon}\C_{2}$ and ${}^{\epsilon}\!\!\stackrel{\ast}
{\C}_{2}$ are transformed into each other under action of anti-automorphism
$\cA\longrightarrow\widetilde{\cA^{\star}}$ (this property of neutrino
field known as $CP$-invariance also has place in the framework of algebra
$\C_{3}$).  
\section{Summary}
In conclusion we shall summarize the results obtained above, and also
in previous paper \cite{3}, in the form of following schedule:
\begin{itemize}\item
{\large$$\C_{2}$$}
The simplest Clifford algebra with even dimensionality $\C_{2}$ and
obtained from it under action of anti-automorphism $\cA\longrightarrow
\widetilde{\cA}$ the algebra $\stackrel{\ast}{\C}_{2}$, are described the
photon fields with left-handed and right-handed polarization, respectively.
The algebras $\C_{2}$ and $\stackrel{\ast}{\C}_{2}$ are transformed into
each other under action of anti-automorphism $\cA\longrightarrow
\widetilde{\cA}$.
\item
{\large$$\C_{3}$$}
The simplest Clifford algebra with odd dimensionality $\C_{3}\cong\C_{2}
\oplus\!\!\stackrel{\ast}{\C}_{2}$. The factor-algebras ${}^{\epsilon}\C_{2}$
and ${}^{\epsilon}\!\!\stackrel{\ast}{\C}_{2}$, which are obtained in the
result of homomorphic mappings $\C_{3}\longrightarrow\C_{2}$ and
$\C_{3}\longrightarrow\stackrel{\ast}{\C}_{2}$, are described the left-handed
neutrino and right-handed antineutrino fields, respectively. The 
factor-algebras ${}^{\epsilon}\C_{2}$ and ${}^{\epsilon}\!\!\stackrel{\ast}
{\C}_{2}$ are transformed into each other under action of anti-automorphism
$\cA\longrightarrow\widetilde{\cA^{\star}}$.
\item
{\large$$\C_{4}$$}
The algebra $\C_{4}$ is described the electron field, which by force of
isomorphism $\C_{4}\cong\C_{2}\otimes\!\!\stackrel{\ast}{\C}_{4}$ may be
represented as a tensor product of two photon fields. Analogously, the
algebra $\overline{\C}_{4}$ is described the positron field, which obtained
from $\C_{4}$ by means of anti-automorphism $\cA\longrightarrow
\widetilde{\cA^{\star}}$.
\end{itemize}

It is obvious that this schedule may be continued. Moreover, we assume that
in like manner (that is in the terms of Clifford algebras) may be constructed
the full particle systematics. 

\end{document}